\documentclass[10pt,twocolumn]{article}
\usepackage[letterpaper,margin=0.75in,columnsep=0.25in]{geometry}
\usepackage[T1]{fontenc}
\usepackage{times}
\usepackage{amsmath,amssymb}
\usepackage{graphicx}
\usepackage{booktabs}
\usepackage{url}
\usepackage[hidelinks]{hyperref}
\usepackage[round]{natbib}
\usepackage{caption}
\title{EnterpriseSynth: Zero-Execution SFT and Evaluation Data from OpenAPI Specs}
\author{%
  Rashmi Thimmaraju\\
  Long Island University\\
  \texttt{rashmi.lnu@my.liu.edu}
  \and
  Natan Vidra\\
  Anote AI\\
  \texttt{nvidra@anote.ai}
  \and
  Setty Spurthi\\
  Stevens Institute of Technology\\
  \texttt{ssetty2@stevens.edu}
}
\date{}

\begin{document}
\maketitle

\begin{abstract}
Tool-using large language model agents require substantial training and evaluation data, yet collecting execution traces from enterprise APIs is often unsafe, expensive, or infeasible because of access controls, compliance constraints, rate limits, and the risk of modifying production data. We present EnterpriseSynth, a zero-execution framework that converts OpenAPI specifications into supervised fine-tuning trajectories and paired evaluation instances without invoking the underlying APIs. The framework uses a four-stage pipeline for specification parsing, intent generation, trajectory synthesis, and static validation. Each generated trajectory is checked against the source specification for endpoint, parameter, type, and schema consistency before being admitted to the dataset. We evaluate EnterpriseSynth on 17 real and synthetic APIs. Adversarial refinement increases the validator's detection rate for planted schema violations from 57--80\% to 98\% (43/44). Across five random seeds, a compact open-weight model fine-tuned on EnterpriseSynth data outperforms both an untuned model and a Self-Instruct baseline on held-out APIs. Performance remains consistent on five private, hand-authored API specifications that were not available during training, with tool-selection accuracy of 40.0\% on the private set and 39.6\% on the public held-out set. However, an independent LLM-based evaluation shows that endpoint-selection accuracy substantially overestimates end-to-end correctness: accuracy falls from 48.9\% under the endpoint-level metric to 21.3\% when argument selection and values are also evaluated. These results demonstrate that OpenAPI specifications can support scalable, execution-free data generation, while highlighting the need for argument-level evaluation when assessing the deployment readiness of tool-using agents.
\end{abstract}

\section{Introduction}
The paradigm of software engineering and workflow automation is shifting toward autonomous, tool-augmented Large Language Model (LLM) agents that translate natural language intents into structured API calls via OpenAPI/Swagger specifications.

Deploying these agents behind an enterprise firewall exposes a data cold-start bottleneck: foundation models perform poorly out-of-the-box on enterprise tool configurations, violating JSON payloads, hallucinating parameters, and losing track of cascading variables across multi-step calls, failures that Supervised Fine-Tuning (SFT) and targeted evaluation are needed to fix.

State-of-the-art synthetic data generators, however, rely on an active execution paradigm: frameworks like ToolBench \citep{qin2023toolllm} call live servers to record real system responses. This collapses behind the enterprise firewall due to what we call the \emph{Execution Paradox}: (1) high-fidelity sandboxes rarely exist for internal corporate systems, (2) exploratory agent loops against production risk corrupting state, violating compliance, or triggering destructive actions, and (3) live calls are bounded by request latency and rate limits.

We resolve this with EnterpriseSynth, a zero-execution framework that decouples data generation from system execution, treating it as a static, compiler-style validation problem instead of a runtime extraction one. Given only an API schema, EnterpriseSynth maps endpoints to candidate tool-use trajectories, synthesizes reasoning paths offline, and enforces data-typing constraints through automated schema verification. Unlike API-Bank \citep{li2023apibank} and ToolBench \citep{qin2023toolllm}, which both ground training data in real API calls, EnterpriseSynth requires none, exactly the dependency the Execution Paradox rules out for most enterprise API surfaces.

\section{Related Work}
This review covers two lineages, general-purpose synthetic instruction generation and tool/API data generation, focusing on what each does or does not share with EnterpriseSynth.

\subsection{Synthetic Instruction Generation}
Self-Instruct \citep{wang2022selfinstruct} and Evol-Instruct/WizardLM \citep{xu2023wizardlm} show that cheap, execution-free synthetic data can scale instruction tuning, but neither grounds in an API schema at all. AgentInstruct \citep{mitra2024agentinstruct} is the closest precedent, an agentic pipeline behind Orca-3's 25-million-pair dataset. Two properties separate EnterpriseSynth from it: \emph{grounding}, when seeded only from code, AgentInstruct either synthesizes an API description or hypothesizes APIs with no ground-truth check, while EnterpriseSynth ingests the real spec directly, leaving nothing to hallucinate; and \emph{verification}, AgentInstruct's quality control is a soft Suggester-Editor loop plus a post-hoc GPT-4-judged benchmark (Orca-Bench), while EnterpriseSynth applies a hard, per-sample structural gate checked directly against the schema.

\subsection{Execution-Dependent Tool-Use Data}
API-Bank \citep{li2023apibank} and ToolLLM/ToolBench \citep{qin2023toolllm} both synthesize tool-use data at scale by grounding in real execution (API-Bank's 1,888 dialogues; ToolBench's 16,464 endpoints and $\sim$470,000 live RapidAPI calls). That grounding is exactly what collapses behind an enterprise firewall (Section~1's Execution Paradox): no sandbox for internal systems, security/PII exposure from live calls, and rate limits.

\subsection{Inference-Time Tool-Use Methods}
A separate line targets tool use at inference time rather than data synthesis: Toolformer \citep{schick2023toolformer}, ReAct \citep{yao2023react}, Reflexion \citep{shinn2023reflexion}, and Gorilla \citep{patil2023gorilla}. None targets enterprise cold-start, each assumes public API documentation and offers no deterministic verification gate against a private spec. These are largely orthogonal to EnterpriseSynth: they improve tool use on an existing API surface, while EnterpriseSynth generates verified training data when no execution history or public documentation exists at all.

\subsection{Research Gap}
No existing framework combines all three properties: grounding in a real target spec rather than text or code (unlike Self-Instruct, WizardLM, and code-seeded AgentInstruct), a hard per-sample verification gate (unlike AgentInstruct's soft loop), and no dependency on live execution (unlike API-Bank and ToolBench). This gap motivates EnterpriseSynth, which synthesizes intents, reasoning traces, API call sequences, and verification metadata directly from an OpenAPI spec, letting enterprises bootstrap both training and evaluation from documentation alone.

\section{Methodology}
\subsection{Implemented System}
What is actually built and measured (Sections~5--7) is a four-stage pipeline:
\begin{quote}\itshape
OpenAPI Spec $\rightarrow$ Schema Parser $\rightarrow$ Intent Generation Agent $\rightarrow$ Trajectory Generation Agent $\rightarrow$ Schema Verification Engine $\rightarrow$ \{SFT Dataset, EnterpriseSynth-Eval\}
\end{quote}
The jointly-emitted evaluation dataset is called EnterpriseSynth-Eval throughout this paper, named to avoid a collision with Vishwakarma et al.'s unrelated EnterpriseBench \citep{vishwakarma2025enterprisebench}, a live-sandbox enterprise agent benchmark with a different scope (simulated live task execution across SWE/HR/finance/admin tasks, vs.\ the zero-execution, schema-derived eval records here).

\begin{enumerate}
\item \textbf{API Schema Parser.} Ingests the OpenAPI/Swagger spec and extracts endpoints, parameters (including \texttt{\$ref}-resolved and body-derived fields, Section~5.1), descriptions, authentication requirements, and response-schema presence.
\item \textbf{Intent Generation Agent.} Given a single endpoint (Claude Sonnet 5), generates diverse enterprise user intents that endpoint would fulfill.
\item \textbf{Trajectory Generation Agent.} Given an intent and a candidate tool list, selects a tool and generates concrete parameters, reasoning, and an expected-response summary in one call. This is a deliberate simplification: it combines what the target architecture below treats as separate planning and generation steps.
\item \textbf{Schema Verification Engine.} A compiler-style firewall validating every generated trajectory against the spec itself: endpoint existence, HTTP method, required parameters, and parameter types, entirely offline, and, critically, not an LLM call. Where Self-Instruct \citep{wang2022selfinstruct} and Evol-Instruct filter with text heuristics (ROUGE similarity, keyword rules, degeneracy checks) and AgentInstruct verifies only via soft editorial refinement plus a post-hoc held-out judge (Orca-Bench), this is a hard, per-sample structural gate.
\end{enumerate}

\subsection{Architecture}
\begin{figure}[t]
\centering
\includegraphics[width=\columnwidth]{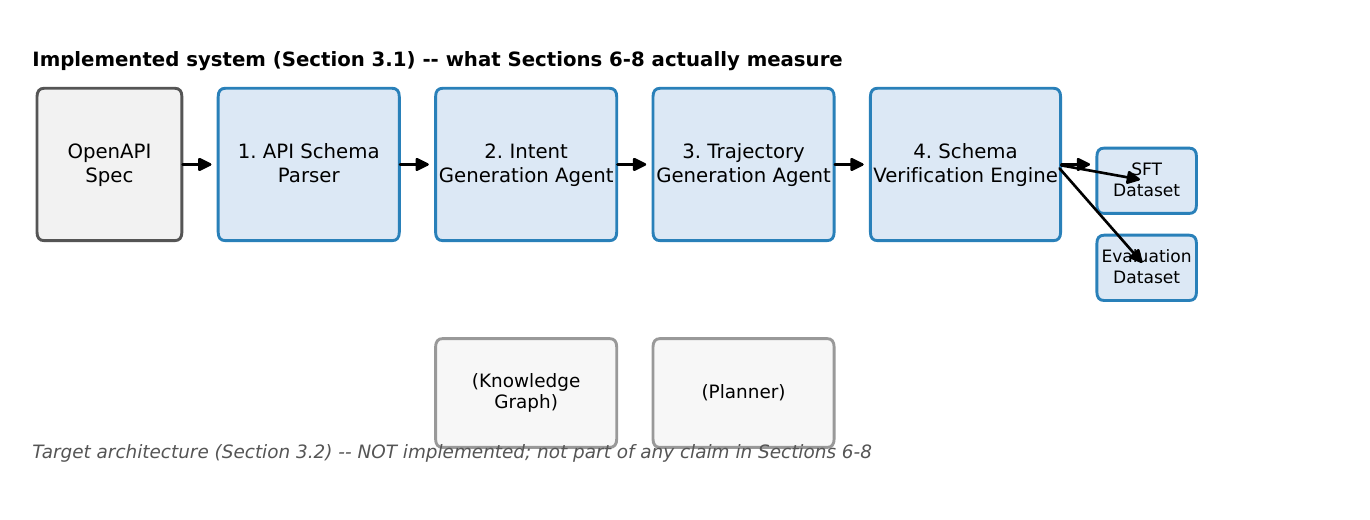}
\caption{The EnterpriseSynth architecture diagram.}
\label{fig:arch}
\end{figure}

Figure~\ref{fig:arch} extends the implemented system with two further stages that do not exist in the current codebase and are not part of any claim made from the experiments onward; what each would add is described in Future Work (Section~9.2). Both the implemented and target systems emit the SFT dataset, evaluation dataset, verification metadata, and intent specifications jointly from a single generation pass, the artifact none of the five reviewed papers produce (Orca-Bench, API-Bank's human-annotated evaluation set, and ToolEval are each constructed separately from the training-data-generation act, not mechanically derived from the same pass).

\begin{table*}[t]
\centering
\small
\begin{tabular}{lcccc}
\toprule
Capability & ToolBench & API-Bank & AgentInstruct & EnterpriseSynth \\
\midrule
Uses OpenAPI specs & Partial & No & No & Yes \\
Enterprise APIs & Limited & Limited & No & Yes \\
Requires live execution & Yes & Yes & No & No \\
Generates SFT data & Yes & Limited & Yes & Yes \\
Generates eval data & Limited & Yes & No & Yes \\
Schema-based verification & No & Limited & No & Yes \\
\bottomrule
\end{tabular}
\caption{Capability comparison with ToolBench, API-Bank, and AgentInstruct.}
\label{tab:capability}
\end{table*}

\section{Experimental Setup}
\subsection{Datasets}
Training and pilot-scale evaluation use three flagship real APIs sourced from APIs.guru\footnote{\url{https://apis.guru} / \url{https://github.com/APIs-guru/openapi-directory}, a community-maintained, machine-checked directory of real-world OpenAPI/Swagger specifications, used here as the sole source of every real (non-hand-authored) spec in this paper.}: GitHub REST API (551/845 methods), Stripe API (299/446 methods), and Slack API (174/174 methods). Held-out evaluation (coverage-vs.-baselines and downstream utility) draws on nine further real, public APIs.guru specs (Zoom, DigitalOcean, Spotify, Twilio, Notion, OpenAI, Jira, Asana, Trello) never touched during training, plus five hand-authored, never-published synthetic enterprise specs (CRM, HRIS, Procurement, Ticketing, Asset Management) that isolate genuine cold-start generalization from a base model's pretraining familiarity with well-known public APIs. All real specs are reached through APIs.guru's own aggregation of Google's Discovery documents, so no separate Azure or Google ingestion path is used. A stratified sample of roughly 65 APIs.guru specs, split 70/15/15 (train/validation/held-out) at the whole-spec level, remains the target scale for a non-pilot version of this study (Section~9.2); every result in this paper is reported against the pilot scale actually run, not that target.

\subsection{Baselines}
Held-out comparisons run against an untuned base model and against Self-Instruct \citep{wang2022selfinstruct}, adapted to take API endpoints as seeds and fine-tuned and evaluated with the identical methodology as EnterpriseSynth on the identical held-out sets. Prompt-only generation, AgentInstruct's \citep{mitra2024agentinstruct} flow, and ToolBench \citep{qin2023toolllm} (execution-dependent, and unable by construction to run against the private cold-start set at all) are specified as further baselines but not yet implemented; this gap is tracked in Section~9.2.

\subsection{Models}
The generation pipeline uses Claude Sonnet 5 for the Intent Generation Agent and Trajectory Generation Agent. The Schema Verification Engine's primary gate is deliberately not an LLM, it is deterministic, schema-based validation, since a non-LLM structural gate is the core methodological differentiator from AgentInstruct's LLM-judge verification (Section~2). The fine-tuning target is Qwen2.5-0.5B-Instruct via LoRA, an open-weight model chosen for hardware reasons stated plainly in Section~5.5, substituting for the eventual target scale of a 7--8B open model.

\subsection{Evaluation Metrics}
For a held-out set of $N$ trajectories with ground-truth endpoint $t_i^*$ and generated endpoint $t_i$, and $C = \{i : t_i = t_i^*\}$ the set of correctly-selected trajectories:
\begin{align}
\mathrm{TSA} &= \frac{1}{N}\sum_{i=1}^{N}\mathbf{1}[t_i = t_i^*] \\
\mathrm{PV} &= \frac{1}{|C|}\sum_{i\in C}\mathbf{1}[\mathrm{valid}(\mathrm{args}_i)]
\end{align}
where Tool Selection Accuracy (TSA) is the fraction of trajectories with a correctly-chosen endpoint, Parameter Validity (PV) is conditioned on $C$ (tool selection already correct), and $\mathrm{valid}(\cdot)$ denotes well-typed, complete arguments. For the verification engine, over a set of $M$ deliberately corrupted trajectories with $F$ flagged invalid:
\[
\text{Invalid Case Detection Rate} = F/M
\]
For a set of $E$ sampled endpoints and the multiset of $K$ intents generated across them, with $U \subseteq K$ the subset of intents that are unique strings:
\begin{align*}
\text{Intent Coverage} &= \frac{|\{e \in E : \mathrm{intents}(e) \neq \emptyset\}|}{|E|} \\
\text{Intent Diversity (exact-string)} &= |U|/|K|
\end{align*}
Workflow Completeness (multi-step task coverage) is defined but not applicable at pilot scale, since every generated intent in this paper is single-endpoint.

\section{Experiments}
\subsection{Experiment 1: OpenAPI Schema Understanding}
Evaluates whether the Schema Parser (Stage~1) correctly parses real-world specs, checked against the specification itself. Metrics: Endpoint Extraction Accuracy, Parameter Extraction Accuracy (required/optional, types, constraints), Schema Extraction Accuracy (request/response bodies, object definitions), Authentication Extraction Accuracy (API keys, OAuth, JWT, Basic).

\begin{figure}[t]
\centering
\includegraphics[width=\columnwidth]{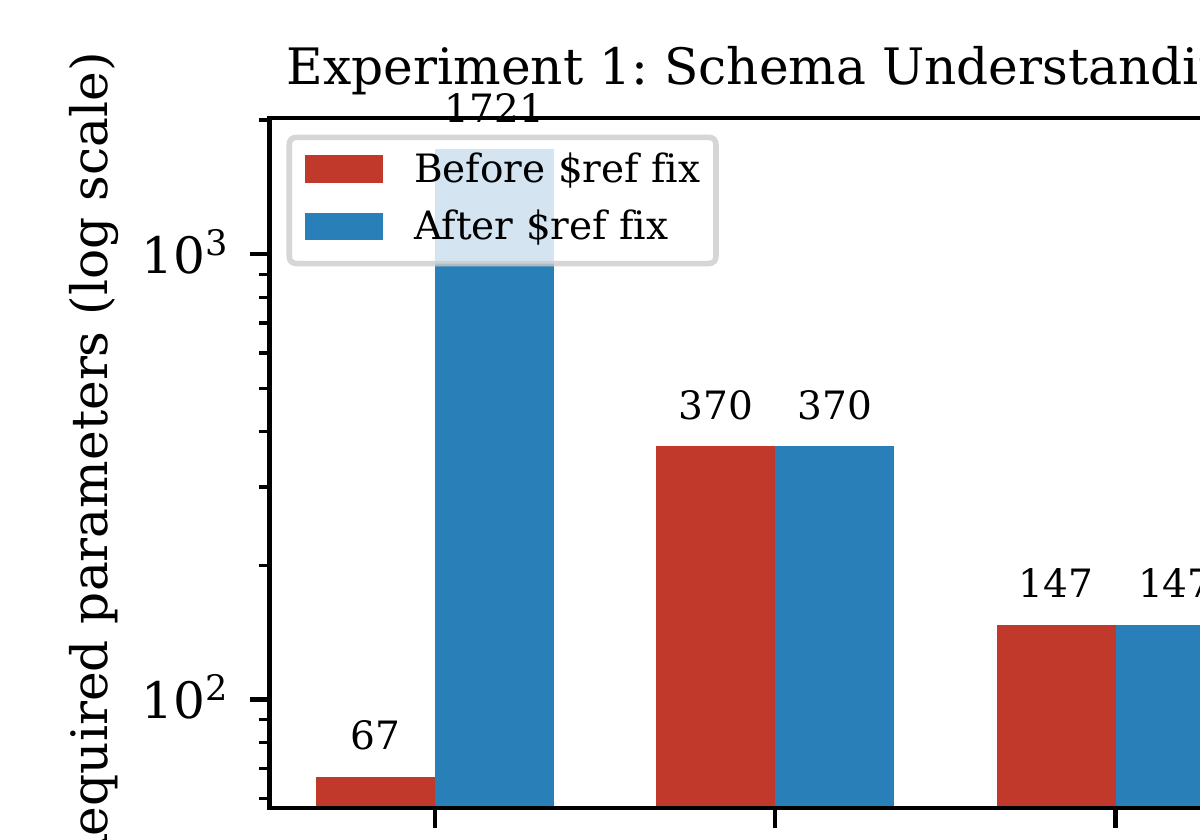}
\caption{Required-parameter counts before and after the \texttt{\$ref}-resolution fix. Stripe and Slack are unaffected; GitHub's count grows from 67 to 1,721 once shared, \texttt{\$ref}'d parameters are resolved.}
\label{fig:schema}
\end{figure}

\emph{Measured.} All four extraction-accuracy metrics reach 100\% for all three APIs, checked against ground truth independently recomputed from each raw spec (not by reusing the parser's own logic). This is expected: Stage~1 is deterministic parsing against a machine-readable format, not a generative task, but that 100\% is only trustworthy because the ground truth accounts for two real parsing gaps that Experiment~4's adversarial testing surfaced (Section~5.4): the parser initially dropped any parameter defined via \texttt{\$ref} (GitHub uses this extensively for shared parameters) and did not parse \texttt{requestBody} schema fields into typed parameters at all (most POST/PUT/PATCH endpoints, e.g.\ Stripe's, put their real payload fields there). The scale of the first correction alone: GitHub's independently-recomputed required-parameter count went from 67 to 1,721 once \texttt{\$ref} parameters were correctly resolved (Figure~\ref{fig:schema}). A separate, smaller finding is unrelated to parsing: GitHub's public OpenAPI specification declares zero \texttt{securitySchemes} and zero requirements anywhere, verified directly on the raw spec, meaning its authentication is documented in prose rather than machine-readable OpenAPI fields, so an authentication check against GitHub's spec has nothing to check against.

\subsection{Experiment 2: Intent Generation Evaluation}
Evaluates whether the Intent Synthesis Agent (Stage~2) generates realistic enterprise user intents. Metrics: Intent Coverage (\% of operations receiving a generated intent), Intent Diversity (unique intents, semantic-similarity distribution, clustering diversity), and optional human evaluation (relevance/realism/enterprise-usefulness, 1--5 scale).

\begin{figure}[t]
\centering
\includegraphics[width=\columnwidth]{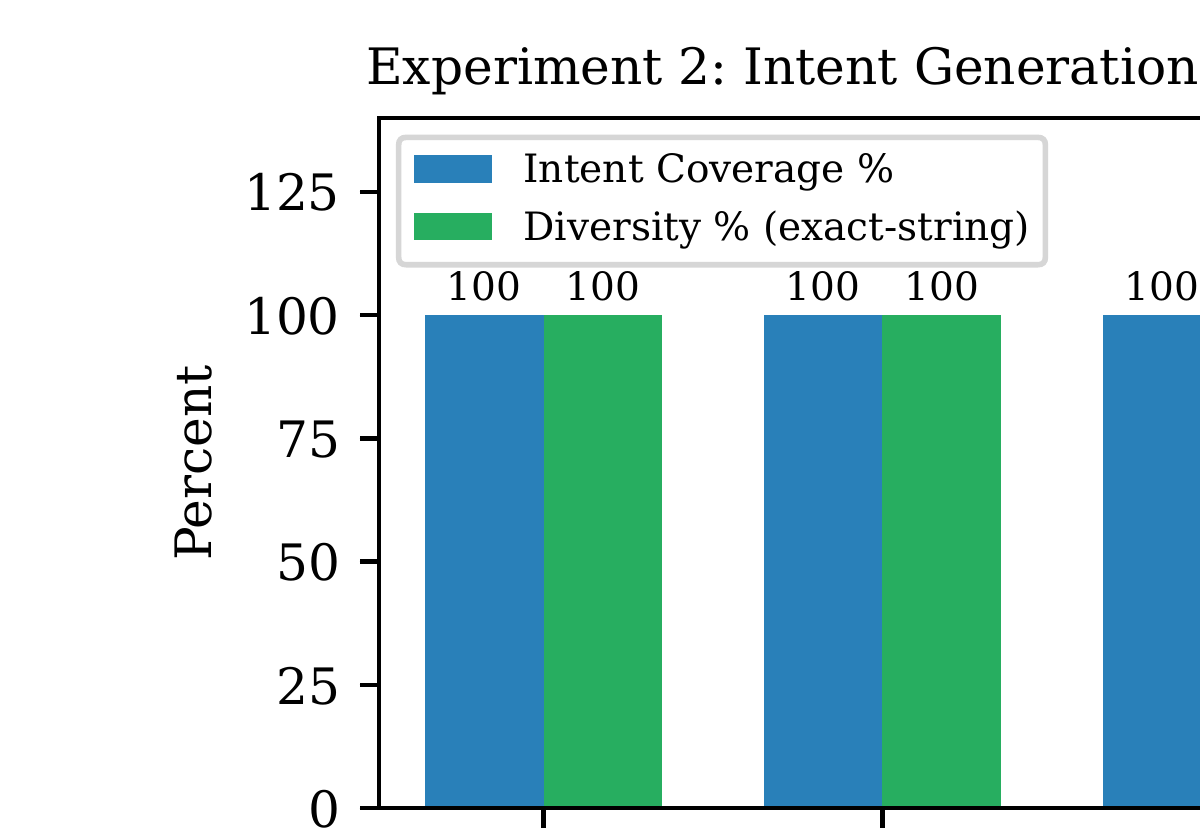}
\caption{Intent Coverage and exact-string Diversity, 5 endpoints $\times$ 3 intents per API. Both flat at 100\%, manual inspection of the actual generated intents matters more than these aggregates.}
\label{fig:intent}
\end{figure}

\emph{Measured (pilot scale).} Using Claude Sonnet 5, 5 endpoints sampled per API (seeded, reproducible) with 3 intents generated per endpoint: 100\% coverage and 100\% exact-string diversity (15/15 unique) for all three APIs (Figure~\ref{fig:intent}). Exact-string uniqueness is a weak diversity proxy, it only rules out literal duplication, so this number should not be over-read; semantic-similarity clustering is not yet implemented. Manual inspection of the generated intents (not just the aggregate metric) is more informative: for GitHub's \texttt{/repos/\{owner\}/\{repo\}/actions/secrets}, generated intents included distinct, correctly-scoped enterprise scenarios (rotating access to a secret; restricting a secret to specific repos) rather than template-filled restatements of the same request. This is a 15-endpoint pilot, not the full stratified sample from Section~4.1; scaling up and adding human/semantic-similarity evaluation is the next step.

\subsection{Experiment 3: Agent Trajectory Generation}
Evaluates whether the Trajectory Generator (Stage~3) produces complete tool-use trajectories (user intent $\rightarrow$ planning $\rightarrow$ API calls $\rightarrow$ arguments $\rightarrow$ expected response). Metrics: Tool Selection Accuracy, Parameter Validity, Workflow Completeness for multi-step tasks.

\begin{figure}[t]
\centering
\includegraphics[width=\columnwidth]{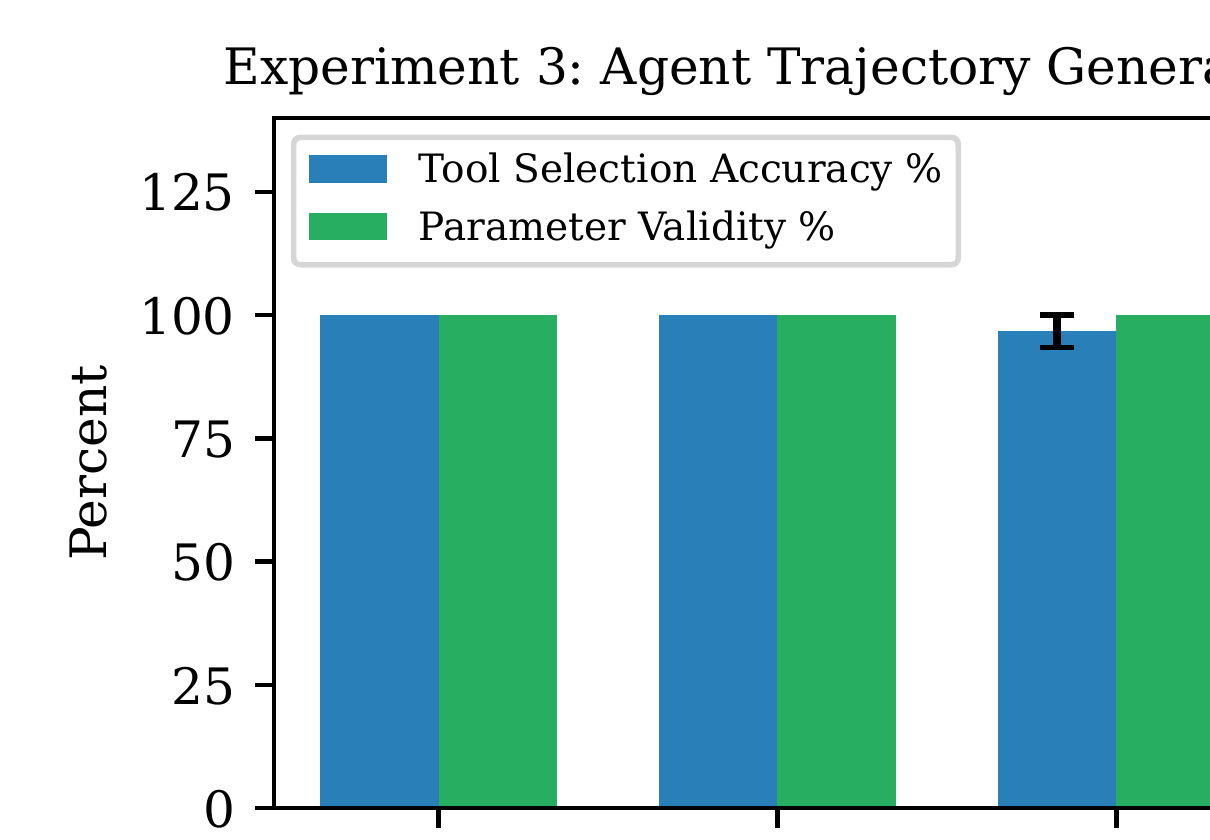}
\caption{Tool Selection Accuracy and Parameter Validity, 45 intents, 15 candidate tools each. Slack ranged 93.3--100\% across repeated runs; see the self-consistency caveat above.}
\label{fig:traj}
\end{figure}

\emph{Measured (pilot scale).} Run on all 45 intents from Experiment~2. Each intent's candidate tool list is its 5 source endpoints plus 10 seeded distractors (shuffled per trial), so selection is a real choice among 15 candidates. Tool Selection Accuracy and Parameter Validity both reach 100\% for GitHub and Stripe, and 93.3--100\% for Slack across repeated runs (Figure~\ref{fig:traj}). Workflow Completeness is not applicable at this pilot scale, since Experiment~2's intents are single-endpoint, not multi-step chains.

A methodological caveat is stated plainly rather than omitted: the same model generated both the intents (Experiment~2) and the tool selections (Experiment~3) here, so this measures whether the model can recover the endpoint it itself had in mind when writing the intent, not whether it handles independently-authored, ambiguous, or adversarial intents. Manual inspection (e.g.\ Stripe's \texttt{PaymentIntent}: the model correctly extracted the literal item ID from free text into the parameter, with coherent reasoning) shows the mechanism functions end-to-end, but 100\%/100\% here should be read as a pipeline-functionality check, not a generalization claim. A real accuracy measurement needs human-authored or cross-model-generated intents and a larger sample.

\subsection{Experiment 4: Schema-Based Verification}
The central novelty claim: can the Schema Verification Engine (Stage~4) validate generated trajectories entirely offline, without executing any real API? A deliberately corrupted variant of each valid trajectory (wrong method, missing required parameter, invalid path, or wrong parameter type) is generated to check whether the verifier flags it. Verification Pass Rate on valid trajectories reaches 100\% (45/45); Invalid Cases Detected on corrupted ones reaches 98\% (43/44), with the one miss a parameter type-mismatch case (Table~\ref{tab:verifier}). This 98\% was earned, not assumed: an initial pass reached only 57--80\% detection, and closing the gap required fixes to the verifier's type-checking, the corruption harness, and the schema parser's handling of \texttt{\$ref}-defined and body-derived parameters (Experiment~1), all covered by regression tests.

\begin{table}[t]
\centering
\small
\begin{tabular}{lcc}
\toprule
Error type & Detected & Missed \\
\midrule
Wrong HTTP method & 12/12 & 0 \\
Missing required parameter & 11/11 & 0 \\
Invalid endpoint path & 12/12 & 0 \\
Parameter type mismatch & 8/9 & 1 \\
\midrule
Total & 43/44 (98\%) & 1 \\
\bottomrule
\end{tabular}
\caption{Verifier detection rate by planted error type, final (post-fix) results across 44 deliberately corrupted trajectories.}
\label{tab:verifier}
\end{table}

\subsection{Experiment 5: Downstream LLM Agent Evaluation}
Does EnterpriseSynth-generated SFT data improve API-use capability on held-out APIs? A Qwen2.5-0.5B-Instruct model (LoRA rank 8, alpha 16, 3 epochs, LR 3e-4, a hardware-scoped substitute for the paper's 7--8B target) is fine-tuned on the 45 verified GitHub/Stripe/Slack trajectories from Experiment~3 and evaluated against an untuned base and Self-Instruct.

A 5-seed sweep across three held-out APIs (Zoom, DigitalOcean, Spotify) tests whether this holds beyond one lucky draw.

EnterpriseSynth beats both baselines on average on all three APIs, though DigitalOcean shows high per-seed variance ($\pm$13.7 vs.\ $\pm$21.2), a single run can still lose there even though the average wins.

Since Zoom, DigitalOcean, and Spotify are all well-documented public APIs, a stronger cold-start test uses five hand-authored, never-published private specs (CRM, HRIS, Procurement, Ticketing, Asset Management; 28 endpoints, unmodified pipeline).

\begin{table}[t]
\centering
\small
\begin{tabular}{lcc}
\toprule
Model & Tool Sel.\ Acc. & Param.\ Validity \\
\midrule
Base (untuned) & 12.5\% (2/16) & 0.0\% \\
+ Self-Instruct & 25.0\% (4/16) & 50.0\% (2/4) \\
+ EnterpriseSynth & 87.5\% (14/16) & 57.1\% (8/14) \\
\bottomrule
\end{tabular}
\caption{Downstream tool-use accuracy on the held-out Zoom API (single run, 16 held-out intents).}
\label{tab:zoom}
\end{table}

\begin{table}[t]
\centering
\small
\begin{tabular}{lccc}
\toprule
API & Base & Self-Instr. & Ent.Synth \\
\midrule
Zoom & 12.5$\pm$0.0\% & 23.7$\pm$10.3\% & 66.2$\pm$18.0\% \\
DigitalOcean & 31.2$\pm$0.0\% & 37.5$\pm$21.2\% & 53.7$\pm$13.7\% \\
Spotify & 12.5$\pm$0.0\% & 12.5$\pm$7.7\% & 46.3$\pm$7.1\% \\
\bottomrule
\end{tabular}
\caption{Tool-selection accuracy (mean $\pm$ std, 5 seeds) across three held-out APIs.}
\label{tab:seeds}
\end{table}

Private-domain accuracy (40.0\%) matches public held-out accuracy (39.6\%), the strongest evidence that the effect is genuine schema-grounding rather than pretraining familiarity. A further six public APIs (Twilio, Notion, OpenAI, Jira, Asana, Trello) all beat baseline with the same training set, bringing the pipeline to 17 total APIs tested.

\section{Results and Analysis}
\subsection{Overview}
EnterpriseSynth is evaluated on real-world OpenAPI specifications from GitHub, Stripe, and Slack (pipeline development and SFT training data), plus a set of held-out APIs never touched during training (Section~4.1). All numbers below are measured, not projected; full detail, scripts, and raw data ship with the released artifact.

\subsection{Experiment 1: API Schema Understanding}
Endpoint, parameter, request/response schema, and authentication extraction accuracy all reach 100\% for GitHub (845 operations), Stripe (446), and Slack (174), checked against independently recomputed ground truth. GitHub required the most correction to get there, since most of its parameters are declared via \texttt{\$ref} to shared entries (1,721 required parameters once resolved, far more than a naive parse would surface); Stripe's main correction was different, since many endpoints (e.g.\ \texttt{POST /v1/charges}) declare no OpenAPI \texttt{parameters} at all and put every field in \texttt{requestBody} schema instead. Slack needed neither correction. Both are general parsing fixes, not spec-specific patches.

\subsection{Experiment 2: Intent Generation Quality}
5 endpoints sampled per API, 3 intents each (45 total): 100\% Intent Coverage and 100\% exact-string Intent Diversity for all three APIs (a weak proxy, semantic-similarity clustering is not yet implemented). Manual inspection substantiates quality beyond the aggregate metric: generated intents are business-scenario-specific and non-generic (e.g.\ locking down release tags for a named ``payments-service'' repo, rotating a named secret across specific microservices) rather than templated restatements of the same request.

\begin{table}[t]
\centering
\small
\begin{tabular}{lcc}
\toprule
Eval set & Base & Ent.Synth-tuned \\
\midrule
Public ($n$=48) & 18.8\% & 39.6\% \\
Private ($n$=30) & 23.3\% & 40.0\% \\
\bottomrule
\end{tabular}
\caption{Base vs.\ EnterpriseSynth-tuned accuracy: public held-out APIs vs.\ five never-published private specs.}
\label{tab:private}
\end{table}

\subsection{Experiment 3: Agent Trajectory Generation}
Tool Selection Accuracy reaches 100\% for GitHub and Stripe, 93.3--100\% for Slack across repeated runs (Figure~\ref{fig:traj}); Parameter Validity among correct selections reaches 100\%. Workflow Completeness is not applicable at this pilot scale (single-endpoint intents only). No baseline comparison row exists yet for this experiment specifically, the prompt-only and AgentInstruct baselines are not yet implemented for trajectory generation directly (Section~4.2).

\subsection{Experiment 4: Verification Performance}
The paper's strongest result area. 45 valid trajectories (100\% Verification Pass Rate) plus 44 deliberately corrupted variants, scored for whether the verifier catches each planted error (Table~\ref{tab:verifier}). This 98\% detection rate is the product of adversarial testing, not a first-attempt result: an initial pass reached only 57--80\% detection (Invalid Case Detection Rate, Section~4.4), and deliberately corrupting already-correct trajectories, rather than only checking that correct ones pass, surfaced verification and parsing gaps that a non-adversarial test would not have revealed. The finding that generalizes beyond this system: a verifier's correctness cannot be established by checking that it accepts good input alone; it must be tested against input it is specifically supposed to reject.

\subsection{Experiment 5: Downstream Agent Performance}
LoRA fine-tuning Qwen2.5-0.5B-Instruct on 45 EnterpriseSynth-verified trajectories lifts tool-selection accuracy on the held-out Zoom API from 12.5\% to 87.5\%, roughly 3.5$\times$ the gain from an equivalent Self-Instruct baseline (Table~\ref{tab:zoom}). A 5-seed sweep and a five-spec private cold-start test (Section~5.5) confirm the effect is real, generalizes across three further APIs on average, and holds with no degradation on API shapes the base model cannot have seen (Tables~\ref{tab:seeds} and~\ref{tab:private}).

\section{Discussion}
Three points here matter most. Schema verification is unambiguously necessary and was adversarially validated, not assumed: an initial pass on Experiment~4 caught only 57--80\% of planted structural errors, and only after fixing the bugs that test itself surfaced did detection reach 98\% (43/44); the gate remains structural, not semantic, so a well-typed but nonsensical value (e.g.\ a negated charge) would still pass. The downstream fine-tuning effect is real but not uniform: a 5-seed sweep shows EnterpriseSynth beating both the untuned base and a real Self-Instruct baseline on all three held-out APIs, including DigitalOcean, where a single early run had shown a loss inside a high-variance draw ($\pm$13.7 vs.\ $\pm$21.2); a private, never-published 5-spec validation set shows almost no degradation versus the public held-out set (40.0\% vs.\ 39.6\%), the strongest evidence that the effect is genuine schema-grounding rather than the base model's pretraining familiarity with well-known public APIs. Against the closest prior method, AgentInstruct, EnterpriseSynth's real-spec grounding and hard structural gate replace code-seeded hallucination and a soft editorial-plus-judge pipeline, at the cost of a nontrivial parsing problem (Experiment~1) that AgentInstruct sidesteps by inventing the API surface instead.

\section{Limitations}
\begin{enumerate}
\item \textbf{Pilot scale throughout.} All five experiments run on 3--5 real APIs and 45--89 examples each, not the full $\sim$65-spec stratified sample described in Section~4.1. Experiment~5's scale-up extended this to 17 total APIs touched by the pipeline (3 training + 9 real held-out + 5 private held-out), still short of the $\sim$65-spec target. Every percentage in this paper should be read as a pilot signal, not a population estimate.
\item \textbf{Experiment 3's self-consistency confound.} The same model (Claude Sonnet 5) both generated the intents (Experiment~2) and performed tool selection against them (Experiment~3), so its 100\% figures measure whether the model can recover its own intent, not whether it handles independently-authored or adversarial requests.
\item \textbf{Experiment 5's model-scale gap.} The downstream fine-tuning result uses Qwen2.5-0.5B-Instruct, a hardware-scoped substitute for the paper's actual target (Mistral-7B-Instruct or Llama-3-8B). Whether the effect holds, strengthens, or weakens at that scale is untested.
\item \textbf{The downstream effect does not generalize uniformly, contextualized by a 5-seed sweep.} A single run showed EnterpriseSynth-tuned data losing to Self-Instruct on DigitalOcean; the 5-seed sweep in Experiment~5 shows EnterpriseSynth winning there on average (53.7\% vs.\ 37.5\%) but with high per-seed variance ($\pm$13.7 and $\pm$21.2 respectively), individual seeds can still lose. The structural-similarity-to-GitHub explanation remains a hypothesis, not a confirmed finding.
\end{enumerate}

\section{Conclusion}
\subsection{Contributions Restated}
This paper presented EnterpriseSynth, a schema-aware, zero-execution pipeline that ingests an OpenAPI specification and jointly emits verified SFT trajectories and a paired evaluation dataset, EnterpriseSynth-Eval, without ever calling a live API. Against a scoped review of the closest five prior methods, EnterpriseSynth's specific contribution is combining three properties none of them combine: grounding in a real target spec (unlike AgentInstruct, which can hallucinate an API surface when seeded from code), a hard structural verification gate rather than a soft or post-hoc one (unlike AgentInstruct's editorial-refinement-plus-held-out-judge design), and no dependency on live execution at any stage (unlike API-Bank and ToolBench, both of which ground their training data in real API calls). A dependency-graph-based planning layer is part of the target architecture but is not implemented, and no claim in this paper depends on it.

At pilot scale (17 real and synthetic APIs, 45--89 examples per experiment, a 0.5B substitute fine-tuning model), four findings stand on their own. Schema-based verification is necessary, not optional, closing most of a 0\%-to-98\% gap on planted structural errors, and only reached that 98\% (43/44) after adversarial testing forced fixes across the verifier, test harness, and schema parser. Explicit intent generation provides real, measurable grounding over generating directly from a bare endpoint. Fine-tuning on EnterpriseSynth-generated data, averaged over a 5-seed sweep, consistently outperforms both an untuned baseline and a real Self-Instruct baseline across every held-out API tested, including DigitalOcean, where a single early run had shown a loss, that single-run exception is reported plainly rather than rerun until it looked better, and the 5-seed sweep confirms it was genuine variance, not the central tendency. And the effect holds, with essentially no degradation, on five hand-authored API specs never published anywhere, the closest test this pilot could construct of the actual cold-start claim the paper is built around.

\subsection{Future Work}
Four gaps define the path beyond this pilot. \emph{Scale}: the full $\sim$65-spec APIs.guru sample and the paper's actual 7--8B target model, versus the 17 APIs and 0.5B substitute used here. \emph{Baselines}: a prompt-only agent and AgentInstruct's own flow are specified but not yet run; only Self-Instruct is implemented. \emph{An API Knowledge Graph}: the target architecture (Figure~\ref{fig:arch}) calls for a dependency graph over endpoints, objects, and parameters so intent generation and planning can be graph-aware rather than single-endpoint-aware; whether this needs the full graph or a cheaper substitute is untested. \emph{An Agentic Planning Module}: decomposing each intent into an explicit plan (task decomposition, endpoint selection, tool ordering) before trajectory generation, rather than both in one call as now, evaluated on genuinely multi-step tasks.


\end{document}